\documentclass[11pt,amsmath,amssymb,nofootinbib,notitlepage,longbibliography]{revtex4-1}
\usepackage{amsmath,amssymb}
\usepackage{slashed}
\usepackage{graphicx}
\usepackage{bm}
\usepackage[top=1.0in,bottom=1.0in,left=1.0in,right=1.0in]{geometry}
\usepackage[colorlinks,linkcolor=blue,citecolor=blue]{hyperref}

\newcommand{\be}{\begin{equation}}
\newcommand{\ee}{\end{equation}}
\newcommand{\bea}{\begin{eqnarray}}
\newcommand{\eea}{\end{eqnarray}}
\newcommand{\ba}{\begin{eqnarray}}
\newcommand{\ea}{\end{eqnarray}}

\begin{document}

\title{Holographic charm and bottom pentaquarks I\\ Mass spectra with spin effects}


\author{Yizhuang Liu}
\email{yizhuang.liu@uj.edu.pl}
\affiliation{Institute of Theoretical Physics,
Jagiellonian University, 30-348 Kraków, Poland}

\author{Maciej A. Nowak}
\email{maciej.a.nowak@uj.edu.pl}
\affiliation{Institute of Theoretical Physics and Mark Kac Center for Complex Systems Research,
Jagiellonian University, 30-348 Kraków, Poland}

\author{Ismail Zahed}
\email{ismail.zahed@stonybrook.edu}
\affiliation{Center for Nuclear Theory, Department of Physics and Astronomy, Stony Brook University, Stony Brook, New York 11794--3800, USA}



\begin{abstract}
We revisit  the three non-strange pentaquarks $[\frac 12\frac 12^-]_{S=0,1}$  and $[\frac 12\frac 32^-]_{S=1}$
 predicted  using the holographic dual description,
where chiral and heavy quark symmetry are manifest in the triple limit
of a large number of colors, large quark mass and strong $^\prime$t Hooft
gauge coupling.   In the heavy quark limit, the pentaquarks  with internal heavy quark spin $S$ are all degenerate.
The holographic pentaquarks are dual to an instanton bound to heavy mesons in bulk, without
 the shortcomings related to the nature of the  interaction and the choice of the
hard core inherent to the molecular constructions. We explicitly derive the
spin-spin  and spin-orbit  couplings  arising from next to leading order in the heavy quark mass, and lift  totally the internal spin degeneray, in  fair
agreement with the newly reported charmed pentaquarks from LHCb. New charm and bottom pentaquark states are predicted.
\end{abstract}

\maketitle

\section{Introduction}

Recently the LHCb collaboration has revisited its analysis of the pentaquark states using a nine-fold increase in reconstructed
$\Lambda_b^0\rightarrow J/\Psi pK^-$ decays from the LHCb Run-2 data batch at 13 TeV~\cite{Aaij:2019vzc}. The LHCb
new high statistics analysis shows that the previously reported $P_c^+(4450)$~\cite{Aaij:2015tga}
 splits into two narrow peaks  $P_c^+(4440)$ and  $P_c^+(4457)$
 below the $\Sigma_c^+\bar D^{*0}$ threshold, with the appearance of a new and narrow $P_c^+(4312)$ state  below the
$\Sigma_c^+\bar D^{0}$. The evidence for the previously reported state $P_c^+(4380)$~\cite{Aaij:2015tga} has
 weakened.

 We regard the new LHCb data  as evidence that supports the three lowest non-strange pentaquarks  with spin-isospin assignments
 $[\frac 12\frac 12^-]_{S=0,1}$  and $[\frac 12\frac 32^-]_{S=1}$  predicted by holography~\cite{Liu:2017xzo}, in the
 triple limit of a large number of colors, strong $^\prime$t Hooft gauge coupling and a large quark mass.  More
 importantly, we will show below that  the degeneracy in the internal heavy quark spin $S=1$,  is lifted by spin-orbit
 effects at next to leading order in the heavy quark mass as heavy quark symmetry is broken, in fair agreement with the
 new data.  Furthermore, we regard the closeness
 of the pentaquarks $P_c^+(4457)$ and $P_c^+(4312)$ to the $\Sigma_c^+ \bar{D}^{*0}$ and $\Sigma_c^+ \bar D^0$ tresholds respectively,
 as further evidence in support of  this construction, as both tresholds coalesce in the heavy quark limit.


Pentaquark states with hidden charm
were initially suggested in~\cite{Wu:2010jy,Karliner:2015ina},
 and since have been  addressed by many
~\cite{Burns:2015dwa,Richard:2016eis,Lebed:2016hpi,Esposito:2016noz,Olsen:2017bmm,Guo:2017jvc,Karliner:2017qhf} (and references therein).
In short, the current descriptions range  from pentaquarks
made of  compact diquarks~\cite{Maiani:2015vwa,Lebed:2015tna}, to hadro-charmonia~\cite{Eides:2019tgv} and  loosely bound hadronic molecules~\cite{Du:2021fmf} (and references therein).
Heavy pentaquarks  as multiquark states  composed of  heavy and light quarks, fall outside the realm of the canonical quark model.
Their description calls  for a novel hadronic paradigm with manifest chiral and heavy quark symmetry.

It is well established that chiral symmetry dictates most of the interactions between  light quarks, while heavy quark symmetry
organizes the spin interactions between heavy quarks~\cite{Shuryak:1981fza,Isgur:1991wq}. Both symmetries are inter-twined by the phenomenon  of chiral
doubling~\cite{Nowak:1992um,Bardeen:1993ae,Nowak:2004jg} as shown experimentally in~\cite{Aubert:2003fg,Besson:2003cp}.
Therefore, a theoretical approach to the multiquark states should have manifest chiral and heavy quark symmetry, a clear organizational
principle in the confining regime, and should address concisely the multi-body bound state problem.

The holographic principle in general~\cite{Maldacena:1997re,Erlich:2005qh},
and the D4-D8-D$\bar 8$ holographic set-up in particular~\cite{Sakai:2004cn} provide a framework for addressing QCD in the
infrared in the double limit of a large number of colors and strong $^\prime$t Hooft gauge coupling $\lambda=g_{YM}^2N_c$. It is
confining and exhibits spontaneous  chiral symmetry breaking geometrically. The light meson sector is well described by an effective action
with manifest chiral symmetry and very few parameters, yet totally
in line with  more elaborate  effective theories of QCD~\cite{Fujiwara:1984mp}. The same set-up can be minimally
modified to account for the description of heavy-light mesons, with manifest  heavy
quark symmetry~\cite{Liu:2016iqo,Liu:2017xzo,Liu:2017frj,Li:2017dml,Fujii:2020jre}.

Light and heavy-light baryons are dual to instantons and instanton-heavy meson bound states in
bulk~\cite{Hata:2007mb,Hashimoto:2008zw,Kim:2008pw,Hata:2007tn,Hashimoto:2009st,Lau:2016dxk}, providing a
robust geometrical  approach to the multi-body bound state problem.
The holographic construction provides a  dual realization of  the chiral soliton approach and its bound states variants~\cite{Zahed:1986qz,Rho:1992yy,Rho:MULTI},
without the shortcomings of the derivative expansion. It is a geometrical  realization of the molecular  approach
~\cite{Wu:2010jy,Karliner:2015ina}, without the ambiguities of the nature of the meson exchanges, and the arbitrariness in the choice of the
many couplings and form factors~\cite{Lin:2019qiv}. Alternative holographic models for the description of heavy hadrons have been
developed  in~\cite{Dosch:2015bca,Sonnenschein:2018fph}.

 The organization of the paper is as follows: in section~\ref{DBI} we recall the essential aspects of the $N_f=2$ heavy light effective
 action  in leading order in the heavy quark mass introduced in~\cite{Liu:2016iqo,Liu:2017xzo}.  In section~\ref{ORDER_MH2} we
 extend this analysis at next to leading order in the heavy quark mass for the bound heavy baryons seeded by  instantons in bulk.
 In section~\ref{SPIN} we detail the spin-orbit and spin-spin effects for the heavy baryons and their exotics. The induced quantum
 effective potentials are made explicit in section~\ref{INDUCED}. In section~\ref{SPECTRA} we derive the holographic mass formula for
 the heavy-light baryons and their exotic pentaquarks including the spin contributions. By  adjusting the chief Kaluza-Klein scale used
 in~\cite{Liu:2017xzo,Liu:2017frj}, a more refined heavy baryon spectrum emerges, including the newly reported charmed pentaquarks by LHCb.
 Our conclusions are in section~\ref{CONCLUSION}.  A number of Appendices are added to support the various results.

\section{ Holographic heavy-light effective action}
\label{DBI}

The D4-D8-D$\bar 8$ set-up for light flavor branes is standard~\cite{Sakai:2004cn}.  The minimal modification that accommodates
heavy mesons makes use of an extra heavy brane as discussed in~\cite{Liu:2016iqo,Liu:2017xzo,Liu:2017frj}. It consists
of $N_f$ light D8-D$\bar 8$ branes  (L) and one heavy (H) probe brane in the cigar-shaped geometry that
spontaneously breaks chiral symmetry.
We assume that the L-brane world volume consists of $R^{4}\times S^1\times S^4$ with
$[0-9]$-dimensions.  The light 8-branes are embedded in the $[0-3+5-9]$-dimensions and set
at the antipodes of $S^1$ which lies in the 4-dimension.
The warped $[5-9]$-space is characterized by a finite size  $R$ and a horizon at $U_{KK}$.


\subsection{Dirac-Born-Infeld (DBI) action}


The effective action on the probe L-branes
consists of the non-Abelian DBI   and Chern-Simons  action.
After integrating over the $S^4$, the leading contribution in $1/\lambda$ to the DBI action is

\bea
\label{1}
S_{\rm DBI}\approx -\kappa\int d^4x dz\,{\rm Tr}\left({\bf f}(z){\bf F}_{\mu\nu}{\bf F}^{\mu\nu}+{\bf g}(z){\bf F}_{\mu z}{\bf F}^{\nu z}\right) \ .
\eea
The warping factors are

\be
{\bf f}(z)=\frac{R^3}{4U_z}\,,\qquad {\bf g}(z)=\frac{9}{8}\frac{U_z^3}{U_{KK}} \ ,
\ee
with $U_z^3=U_{KK}^3+U_{KK}z^2$, and $\kappa\equiv a\lambda N_c$ and
$a=1/(216\pi^3)$~\cite{Sakai:2004cn}. All dimensions  are in units of $M_{KK}$ (Kaluza-Klein scale) unless
given explicitly. Our conventions are $(-1,1,1,1,1)$ with $A_{M}^{\dagger}=-A_M$ and
the labels $M,N$ running over $\mu,z$ only in this section.
The effective fields in the field strengths are~\cite{Liu:2016iqo,Liu:2017xzo}

\bea
\label{2}
{\bf F}_{MN}=\left(\begin{array}{cc}
F_{MN}-\Phi_{[M}\Phi_{N]}^{\dagger}&\partial_{[M}\Phi_{N]}+A_{[M}\Phi_{N]}\\
-\partial_{[M}\Phi^{\dagger}_{N]}-\Phi^{\dagger}_{[M}A_{N]}&-\Phi^{\dagger}_{[M}\Phi_{N]}
\end{array}\right) \ .
\eea
The  matrix valued 1-form gauge field is
\be
\label{7}
{\bf A}=\left(\begin{array}{cc}
A&\Phi\\
-\Phi^{\dagger}&0
\end{array}\right) \ ,
\ee
For $N_f=2$, the naive Chern-Simons 5-form is

\be
\label{CSNAIVE}
S_{CS}=\frac{iN_c}{24\pi^2}\int_{M_5}\,{\rm Tr}\left(AF^2-\frac{1}{2}A^3F+\frac{1}{10}A^5\right) \ .
\ee
We note that for only $N_f>2$ it
fails to reproduce the correct transformation law under the combined gauge and chiral transformations~\cite{Hata:2007tn}.
In particular, when addressing the $N_f=3$ baryon spectra, (\ref{CSNAIVE}) does not reproduce the important hypercharge
constraint~\cite{Hata:2007tn}, but can be minimally modified to do that.

For $N_f$ coincidental branes, the $\Phi$ multiplet is massless, but for separated branes
 they are massive with the additional contribution

\bea
\label{8X3}
\frac 12 m_H^2 {\rm Tr}\left(\Phi^\dagger_M \Phi_M\right) \ .
\eea
The value of $m_H$ is related to the separation between the light and heavy branes,
which is about the length of  the HL string. It is related to the heavy meson masses $M_{D}=1870$ MeV (charmed)
and $M_B=5279$ MeV (bottomed)
through~\cite{Liu:2016iqo}
\be
\label{MDMH}
M_{D,B}=m_{H}+\frac{M_{KK}}{2\sqrt{2}} \ .
\ee
Given $M_{KK}$ and $M_{D,B}$, the mass parameter $m_H$ is therefore totally fixed.



\subsection{Light fields}

In the coincidental brane limit,
light baryons are interchangeably described as a flavor instanton or a D4 brane wrapping the $S^4$.
The instanton mass is $M_0=8\pi^2\kappa$ in units of $M_{KK}$.
The instanton size is small with $\rho\sim 1/\sqrt{\lambda}$ after balancing  the order $\lambda$
bulk gravitational attraction  with the subleading and of order $\lambda^0$ U(1) induced topological
repulsion~\cite{Sakai:2004cn}. The bulk instanton is described by the  O(4) gauge field

\be
\label{XS3}
A_{M}(y)=-\overline{\sigma}_{MN}\partial_NF(y)\qquad \left.F_{zm}(y)\right|_{|y|=R}=0 \ .
\ee
From hereon $M,N$ run only over $1,2,3,z$ unless specified otherwise.
If $\rho\sim 1/\sqrt{\lambda}$
is the  typical size of these tunneling configurations, then it is natural to recast the DBI action using the rescaling

\bea
\label{S3}
(x_0, x_{M})\rightarrow (x_0,x_{M}/\sqrt{\lambda}), \sqrt{\lambda}\rho\rightarrow \rho\qquad\qquad
(A_{0},A_M)\rightarrow (A_0, \sqrt{\lambda}A_M)
\eea
The rescaled fields  satisfy the equations of motion

\be
\label{HL1}
D_{M}F_{MN}=0\qquad \partial_M^2A_0=-\frac 1{32\pi^2 a}F_{aMN}\star {F}_{aMN} \ ,
\ee
with the use of the Hodge dual notation.

\subsection{Heavy-light fields}

Let  $(\Phi_0, \Phi_M)$ be the pair of heavy quantum fields that bind to the tunneling configuration above.
If again  $\rho\sim 1/\sqrt{\lambda}$ is their typical size, then it is natural to recast the heavy-light part of
the DBI action using the additional rescaling

\be
\label{S3S}
(\Phi_0,\Phi_M)\rightarrow (\Phi_0, \sqrt{\lambda}\Phi_M)  \ .
\ee
The interactions between the light gauge fields $(A_0, A_M)$ and
the heavy fields $(\Phi_0,\Phi_M)$  to quadratic order split to several contributions~\cite{Liu:2016iqo,Liu:2017xzo}

\be
\label{RS1}
{\cal L}=aN_c\lambda {\cal L}_{0}+aN_c{\cal L}_{1}+{\cal L}_{CS}  \ .
\ee
which are quoted in (\ref{ACTIONALL}). We start by recalling the leading contributions in $1/m_H$ stemming
from (\ref{RS1}) as thoroughly discussed in~\cite{Liu:2017xzo,Liu:2017frj}.  For that, we split
$\Phi_{M}=\phi_{M}e^{-im_Hx_0}$ for particles ($m_H\rightarrow -m_H$ for anti-particles). The leading
order contribution takes the form

\be
\label{RX66X}
{\cal L}_0=-\frac 12 \left|f_{MN}-\star f_{MN}\right|^2+2\phi_M^\dagger (F_{MN}-\star F_{MN})\phi_N \ ,
\ee
subject to  the constraint equation $D_M\phi_M=0$ with

\be
f_{MN}=\partial_{[M}\phi_{N]}+A_{[M}\phi_{N]} \ ,
\ee
while the subleading contributions in (\ref{RS1}) to order $\lambda^0m_H$  simplify to

\bea
\label{RX5}
\frac{{\cal L}_{1}}{aN_c}\rightarrow 4m_H\phi^{\dagger}_{M}iD_0\phi_{M}\qquad\qquad
{\cal L}_{CS}\rightarrow \frac{m_H N_c}{16\pi^2}\phi^{\dagger}_{M}\star F_{MN}\phi_{N} \ . 
\eea
For self-dual light gauge fields with $F_{MN}=\star F_{MN}$,
the last contribution in (\ref{RX66X}) vanishes, and the minimum is reached for
$f_{MN}=\star f_{MN}$. This observation when combined with the transversality condition for $D_M\phi_M=0$,
amounts to a first order equation for the combination $\psi=\bar \sigma_{M}\phi_{M}$ with $\sigma_M =(i, \vec \sigma)$, i.e.

 \be
 \label{RX66}
 \sigma_{M}D_{M}\psi= D \psi =0   \ ,
 \ee
 as noted in~\cite{Liu:2016iqo,Liu:2017xzo}. In a self-dual gauge configuration, the heavy spin-1 meson transmutes
 to a massless  spin-$\frac 12$ spinor that is BPS bound in leading order.

\section{The order ${1}/{m_H^2}$ Lagrangian}
\label{ORDER_MH2}

To account for the spin effects and the breaking of heavy quark symmetry we need to account for the
$1/m_H$ contributions to (\ref{RS1}-\ref{RX5}). This will be sought by restricting the quantum and
heavy fields to the quantum moduli. More specifically, we choose to parametrize the fields using

\bea
\label{SP1}
&&A_M(t,x)=V(A_M^{cl}-i\partial_M)V^{-1}, \qquad A_0(t,x)=0 \nonumber\\
&&\Phi_M(t,x)=\frac{e^{-im_Ht}}{\sqrt{16m_HaN_c}}V(t,x)f(X(t),Z(t))\bar \sigma_M  \chi(t) \ .
\eea
which is equivalent to

\bea
\label{SP2}
&&A_M(t,x)=A_M^{cl}(X(t),Z(t)), \qquad A_0(t,x)=-iV\partial_tV^{-1}\equiv \Phi  \nonumber\\
&&\Phi_M(t,x)=f(X(t),Z(t))\bar \sigma_M  \chi(t) \ .
\eea
after gauge transformation. The $\Phi$ is parameterized as

\begin{align}
\label{SP3}
\Phi=-\dot X_N A^{\rm cl}_N+\chi^a \Phi_a \ ,
\end{align}
where $\Phi_a$ diagonalizes $D_M^{\rm cl}D_M^{\rm cl}\Phi_a$ and where

\begin{align}
\label{SP4}
\chi^a={\rm tr}(\tau^a {\bf a}^{-1}\dot {\bf a})  \ ,
\end{align}
are expressed in terms of the collective variables ${\bf a} \in {\rm SU(2)}$ for a rigid ${\rm SU(2)}$ rotation.
The temporal component $\Phi_0$ satisfies the constraint

\begin{align}
\label{SP5}
(-D_M^2+m_H^2) \Phi_0+2F_{M0}\Phi_M-\frac{i}{16\pi^2 a}F_{PQ}(\partial_P+A_P) \Phi_Q=0 \ .
\end{align}
and in leading order in $1/m_H$ can be ignored.


Inserting the expansion (\ref{SP2}) in (\ref{ACTIONALL}) yields the quadratic $1/m_H$ contributions,

\begin{align}
\label{SP6}
{\cal L}_{\rm quadratic}=\frac{1}{8m_H}\chi^{\dagger}\sigma_M f(\partial_t-A_0-\Phi)(\partial_t-A_0-\Phi)\bar \sigma_M f \chi-
\tilde {\cal L}_1+\tilde {\cal L}_{\rm CS} \ ,
\end{align}
Here  $\tilde {\cal L}_{\rm CS}$ contains only $\Phi_M$.   Each of the contribution in (\ref{SP6}) is discussed in Appendix~\ref{HLACTION}.
Combining the results (\ref{eq:LCS},\ref{eq:finalwrap},\ref{L0X}) we have for the quadratic contributions to order $\frac{1}{m_H}$

\begin{align}
\label{eq:final1}
{\cal L}_{\rm quadratic}=&\frac{1}{m_H}\bigg(\frac{c_1}{\pi^4a^2\rho^4}\chi^{\dagger}\chi+i\frac{c_2}{\pi^2 a \rho^2} \chi^{\dagger}\dot \chi+i\frac{c_3}{\pi^2 a \rho^2}\chi^{\dagger}\tau^a\chi \chi^a +\frac{\dot \chi^{\dagger}\dot \chi}{2}\bigg)\nonumber\\
&-\frac{37+12\frac{Z^2}{\rho^2}}{192m_H}\chi^{\dagger}\chi
+\left(\frac{1}{4m_H }\frac{\dot\rho^2}{\rho^2}+\frac{ \dot a_I^2}{4m_H}+\frac{\dot X^2}{4m_H \rho^2}\right)\chi^{\dagger}\chi \ ,
\end{align}
with the constants fixed to
\begin{align}
\label{SP7}
    &c_1=\frac{13}{3840}+\frac{7}{1280}=\frac{17}{1920} \ , \nonumber\\
    &c_2=\frac{2}{32} \ , \nonumber\\
    &c_3=\frac{1}{128}+\frac{1}{80}=\frac{13}{640} \ .
\end{align}
In the mean time, one has also to take into account the Chern-Simons term contribution
\begin{align}
\label{SP13}
    -\frac{i}{16\pi^2}\epsilon_{MNPQ}\Phi^{\dagger}_M\Phi_N\Phi^{\dagger}_P\partial_t \Phi_Q+ {\rm c.c.} \ ,
\end{align}
which is
\begin{align}
\label{SP14}
{\cal L}_{\rm quartic, CS}=-\frac{N_c}{5m_Hm_y^2\rho^4}\chi^{\dagger}\tau^a\chi \chi^{\dagger}\tau^a \chi \ ,
\end{align}
where $m_y=16\pi^2a$.

The above analysis ignores the Coulomb back reaction (repulsion from the bound charged fields)
as we discussed in~\cite{Liu:2016iqo,Liu:2017xzo} and can lead to instabilities.
In Appendix~\ref{BACK} we detail the back-reaction from the Coulomb field with the final result
for (\ref{eq:final1}) to order ${\cal O}({1}/{m_H^2})$

\begin{align}\label{SP15}
{\cal L}\rightarrow\,\,& i\chi^{\dagger}\dot \chi+\frac{1}{2m_H}\dot \chi^{\dagger}\dot \chi+\frac{78}{5m_H\tilde \rho^2} i\chi^{\dagger}\tau^a \chi \chi^a-\frac{12}{5m_H\tilde \rho^4}\vec{S}^2 -\frac{37+12\frac{Z^2}{\rho^2}}{192m_H}n \nonumber \\
&+\frac{1}{\tilde \rho^2}\bigg(-\frac{18}{5}+\frac{9}{2}n-\frac{2}{3}n^2\bigg)+\frac{1}{m_H\tilde \rho^4}\bigg(\frac{102}{5}n-\frac{56}{5}n^2+\frac{4}{3}n^3+j\tilde \rho^2\left(\frac{9}{2}-\frac{4n}{3}\right)\bigg) \nonumber \\
&+\frac{1}{m_H^2 \tilde \rho^6}\bigg(-\frac{128 n^4}{45}+\frac{376 n^3}{15}-\frac{4017 n^2}{70}-jn\tilde \rho^2 \left(\frac{56}{5}-\frac{8}{3}n\right)-\frac{2}{3}j^2 \tilde \rho^4 \bigg)   \ ,
\end{align}
with

\begin{align}
j=\frac{i}{2}\left(\chi^{\dagger}\dot\chi-\dot\chi^{\dagger}\chi\right),\  n=\chi^{\dagger}\chi \  ,
\end{align}

This is the first major result of this paper.
We now study the quantization of the (\ref{SP15}) and the ensuing heavy-light
baryonic spectra.

\section{Quantum spin effects}
\label{SPIN}


\subsection{Spin-orbit effect}

The first major spin contribution occurs through the  spin-angular momentum coupling $\chi^a \chi^{\dagger}\tau^a\chi$.
Recall that  $\chi^a$ in modular variables is

\begin{align}
\label{SP16}
    \chi^a=2i(a_4\dot a_a-\dot a_4a_a+\epsilon_{abc}a_b\dot a_c)  \ ,
\end{align}
with $a_4^2+\sum_{a=1}^3a_a^2=1$ parametrizing the $SU(2)\sim S^3$ moduli.
Thus, the canonical momenta for $y_I$ read

\begin{align}
\label{SP17}
    &\Pi_a= m_y\rho^2\dot a_a-\frac{2c_3}{m_H\pi^2a \rho^2}(a_4\chi^{\dagger}\tau_a\chi+\epsilon_{cba}\chi^{\dagger}\tau_c\chi a_b)  \ ,\\
    &\Pi_4=m_y\rho^2\dot a_4+\frac{2c_3}{m_H\pi^2a \rho^2}\chi^{\dagger}\tau_a\chi a_a \ ,
\end{align}
Therefore the spin-orbit contribution to the Hamiltonian is

\begin{align}
\label{SP18}
    H=&\frac{1}{2m_y \rho^2}(-i\nabla_{\rm S^3}+\frac{4c_3}{m_H\pi^2a \rho^2}\vec{r}\times \vec{S})^2 \nonumber \\
    =&\frac{-\nabla^2_{\rm S_3}}{2m_y \rho^2}+\frac{8c_3}{m_H\pi^2a \rho^2}\frac{\vec{L}\cdot \vec{S}}{m_y \rho^2}
    +\bigg(\frac{4c_3}{m_H\pi^2a \rho^2}\bigg)^2\frac{\chi^{\dagger}\frac{\vec \tau}2\chi\cdot\chi^{\dagger}\frac{\vec \tau}2\chi}{2m_y\rho^2} \ ,
\end{align}
with the orbital angular momentum

\begin{align}
\label{SP19}
    \vec{L}=\frac{1}{2}\left(i\vec{a}\nabla_4-ia_4\vec{\nabla}-i\vec{a}\times \vec{\nabla}\right) \ ,
\end{align}
The integer-valued spin of the heavy-light doublet translates to a half-integer spin on the moduli

\begin{align}
\label{SP20}
    \vec{S}=\chi^{\dagger}\frac{\vec \tau}2\chi \ ,
\end{align}
a remarkable transmutation induced by the binding of the zero mode to the instanton in bulk~\cite{Liu:2016iqo}.
To leading order in ${1}/{m_H}$ only the first two contributions in (\ref{SP18}) will be retained.
The last contribution in (\ref{SP18}) is the induced spin-spin  interaction of the heavy mesons and
is suppressed by $1/m_H^2$.

\subsection{Spin effects}

The leading spin effects to order $1/m_H$ stem from the quadratic and quartic $\chi$-contributions detailed above.
The terms with a first order  time-derivative of $\chi$ are

\begin{align}
\label{SP21}
 (1+\frac{3N_c}{2m_Hm_y\rho^2})   i\chi^{\dagger} \dot \chi +\frac{\dot\chi^{\dagger}\dot \chi}{2m_H}-i\frac{2 \chi^{\dagger}\chi(\chi^{\dagger}\dot\chi-\dot \chi^{\dagger}\chi)}{3m_Hm_y\rho^2} \ ,
\end{align}
and imply the equation of motion

\begin{align}
\label{SP22}
    -\frac{1}{2m_H}\partial_t^2\chi+i(1+\frac{3N_c}{2m_Hm_y\rho^2})\partial_t\chi+\frac{3N_c}{2m_y\rho^2}\chi+\frac{4i(\chi^{\dagger}\chi \dot \chi+ \chi^{\dagger}\dot \chi \chi)}{3m_Hm_y\rho^2}=0 \ .
\end{align}
Therefore to second order in ${1}/{m_H}$ one has
\begin{align}
\label{SP23}
    i\partial_t \chi=&-\frac{3 N_c   }{2m_y\rho^2}\chi+\frac{9N_c^2}{8m_Hm_y^2\rho^4}\chi+ \frac{4\chi^{\dagger}\chi \chi N_c}{m_Hm_y^2\rho^4 } \nonumber \\
    +&\frac{81N_c^3}{32m_H^2m_y^3\rho^6}\chi-\frac{32N_c(\chi^{\dagger}\chi)^2\chi}{3m_H^2m_y^2\rho^6}\ ,
\end{align}
from which the Hamiltonian can be easily extracted.

\subsection{Hamiltonian}

With the above in mind and to  obtain the Hamiltonian in leading order in ${1}/{m_H}$, it is sufficient to perform the following substitution

\begin{align}
j \rightarrow -\frac{3N_c}{2 \tilde \rho^2} \ ,
\end{align}
and add
\begin{align}
\delta H=\frac{81}{8m_H \tilde \rho^4} n +\frac{1}{m_H^2 \tilde \rho^6}\left( \frac{81N_c^3}{32}n-\frac{32N_c}{9}n^3\right) \ ,
\end{align}
to the the spin-independent Hamiltonian~\cite{Liu:2016iqo}.

More specifically,  for a single heavy-quark with $n=1$, the total Hamiltonian to order  ${1}/{m_H}$ now reads ($N_c=3$)
\bea
\label{SP25}
    H_{\rm single}&=&\frac{39\vec{L}\cdot \vec{S}}{5m_Hm_y^2(1+\frac{1}{2m_Hm_y\rho^2})\rho^4}+\bigg(-\frac{553}{120m_H m_y^2 \rho^4}+\frac{67.94}{m_H^2 m_y^4 \rho^6}+\frac{37+12\frac{Z^2}{\rho^2}}{192m_H}\bigg) \nonumber \\
    &-&\frac{1}{2m_y \rho^3(1+\frac{1}{2m_Hm_y\rho^2})^2}\frac{\partial}{\partial \rho}\left(\rho^3(1+\frac{1}{2m_Hm_y\rho^2})\frac{\partial}{\partial \rho}\right)+\frac{4\vec{L}^2}{2m_y\rho^2(1+\frac{1}{2m_Hm_y\rho^2})}\ ,\nonumber\\
\eea
The change in the Laplacian is due to the $\frac{\dot\rho^2}{\rho^2}+a_I^2$ term following from the new line element on the moduli

\begin{align}
ds^2=(1+\frac{1}{2m_Hm_y\rho^2})dy_I^2 \ ,
\end{align}
with a change in the  small $\rho$ behavior.
For the penta-quark states where $N_{\bar Q}=N_{Q}=1$, the corresponding Hamiltonian is ($N_c=3$)

\bea
\label{SP26}
    H_{\rm double}&=&\frac{39\vec{L}\cdot \vec{S}}{5m_Hm_y^2(1+\frac{1}{m_Hm_y\rho^2})\rho^4}+\bigg(-\frac{411}{20m_H m_y^2 \rho^4}+\frac{133.30}{m_H^2 m_y^4 \rho^6}+\frac{37+12\frac{Z^2}{\rho^2}}{96m_H}\bigg) \nonumber \\
    &-&\frac{1}{2m_y \rho^3(1+\frac{1}{m_Hm_y\rho^2})^2}\frac{\partial}{\partial \rho}\left(\rho^3(1+\frac{1}{m_Hm_y\rho^2})\frac{\partial}{\partial \rho}\right)+\frac{4\vec{L}^2}{2m_y\rho^2(1+\frac{1}{m_Hm_y\rho^2})} \ .
\eea
Below we solve the corresponding Schroedinger equation numerically.

\section{Induced quantum potentials}
\label{INDUCED}

\subsection{The effective potential for single-heavy quark: $l=0$ state}

For $l=0$, the spin-orbit coupling vanishes, i.e. $\vec{L}\cdot \vec{S}=0$, and  the induced effective potential simplifies to

\begin{align}
\label{SP27}
  V(\rho)=\frac{m_y\omega_\rho^2}{2}\rho^2- \frac{7}{30m_y\rho^2}+\bigg(-\frac{553}{120m_H m_y^2 \rho^4}+\frac{67.94}{m_H^2 m_y^4 \rho^6}\bigg) \ .
\end{align}
Although the sign of the $\frac{1}{\rho^2}$ is negative, the $m_H=\infty$ system is still stable due to the uncertainty principle. Indeed, for small $\rho$, the kinetic contribution is of order
$\frac 1{\rho^2}$ and compensates the negative sign to maintain stability. In this case, the ${1}/{m_H}$ term implies additional repulsion that further stabilizes the system. As $m_H\rightarrow\infty$, the spectrum approaches  the infinite mass limit smoothly.

\subsection{The effective potential for single-heavy quark: $l>0$ state}

For $l=2,4,..$, one has $J=(l\pm 1)/2$. We first consider the $J=(l-1)/2$ case. Again for $N_Q=1$ and $N_c=3$, the effective potential reads

\bea
\label{SP28}
    V\bigg(J=\frac{l-1}{2},\rho\bigg)&=&\frac{1}{2m_y(1+\frac{1}{2m_H m_y\rho^2}) \rho^2}\left(l(l+2)-\frac{(l+2)\alpha N_c}{m_Hm_y\rho^2}+\frac{3\alpha^2N_c^2}{4m_H^2m_y^2\rho^4}\right)\nonumber \\
    &+&\frac{m_y\omega_\rho^2}{2}\rho^2- \frac{7}{30m_y\rho^2}+\bigg(-\frac{553}{120m_H m_y^2 \rho^4}+\frac{67.94}{m_H^2 m_y^4 \rho^6}\bigg)  \ ,
\eea
with $\alpha=\frac{13}{10}$. The ${1}/{m_H^2}$ term due to the spin-orbit coupling is kept to maintain stability at small $\rho$.
The change of the potential as one increases $m_H$ tends to decrease for larger $l$. For $l=2$, the shapes of the potential at $m_H=2$ and $m_H=\infty$ differ moderately, but for $l=2$ the difference is already quite small.

Similarly, in the $J=\frac{l+1}{2}$ case the effective potential is

\bea
\label{SP29}
    V\bigg(J=\frac{l+1}{2},\rho\bigg)&=&\frac{1}{2m_y(1+\frac{1}{2m_Hmy\rho^2}) \rho^2}\left(l(l+2)+\frac{l\alpha N_c}{m_Hm_y\rho^2}+\frac{3\alpha^2N_c^2}{4m_H^2m_y^2\rho^4}\right)\nonumber \\
     &+&\frac{m_y\omega_\rho^2}{2}\rho^2- \frac{7}{30m_y\rho^2}+\bigg(-\frac{553}{120m_H m_y^2 \rho^4}+\frac{67.94}{m_H^2 m_y^4 \rho^6}\bigg)  \ .
\eea
Again, the $1/m_H$ contribution further stabilizes the system and pushes the spectrum a little bit higher.

\subsection{The effective potential for penta-quark state}

Here we focus on the pentaquark states with $N_Q=\bar N_Q=1$ state or hidden $Q=c,b$, with $S=0,1$. For $S=0$, the potential reads
\begin{align}
\label{SP39}
    &V\bigg(J=\frac{l}{2},S=0\bigg)=\frac{m_y\omega_\rho^2\rho^2}{2}+\frac{18}{5m_y\rho^2}+\frac{l(l+2)}{2m_y (1+\frac{1}{m_Hm_y^2\rho^2})\rho^2}+\bigg(-\frac{411}{20m_H m_y^2 \rho^4}+\frac{133.30}{m_H^2 m_y^4 \rho^6}\bigg) \ ,
\end{align}
For $S=1$ we can have  $J=l-1,l,l+1$, and the potential in this case reads

\bea
\label{SP40}
     V(J,S=1,\rho)&=&\frac{1}{2m_y (1+\frac{1}{m_Hm_y^2\rho^2})\rho^2}\left(l(l+2)+\frac{2N_c\Delta(J)\beta}{m_Hm_y\rho^2}+\frac{2N_c^2\beta^2}{m_H^2m_y^2\rho^4}\right)+\frac{18}{5m_y\rho^2}\nonumber \\
     &+&\frac{m_y\omega_\rho^2}{2}\rho^2+\bigg(-\frac{411}{20m_H m_y^2 \rho^4}+\frac{133.30}{m_H^2 m_y^4 \rho^6}\bigg)\ .
\eea
with  $\beta=\frac{13}{10}$ and

\begin{align}
\label{SP41}
    \Delta(J)=J(J+1)-\frac{l(l+2)}{4}-2 \ .
\end{align}
More specifically, for $l=1$ we have $\Delta(J=1/2)=-2$ and $\Delta (J=3/2)=1$.

\begin{figure}[h!]
\includegraphics[width=0.8\columnwidth]{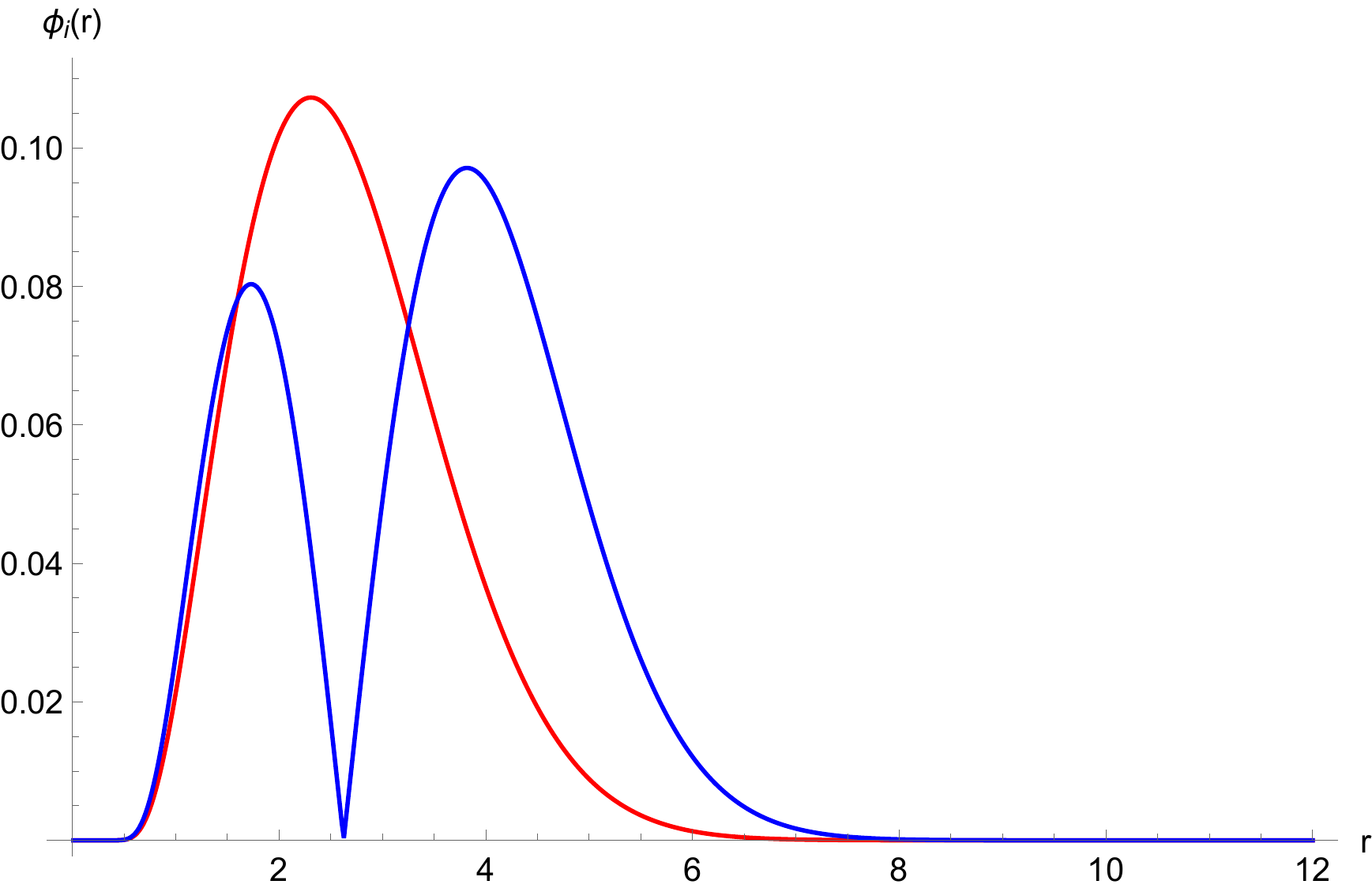}
\includegraphics[width=0.8\columnwidth]{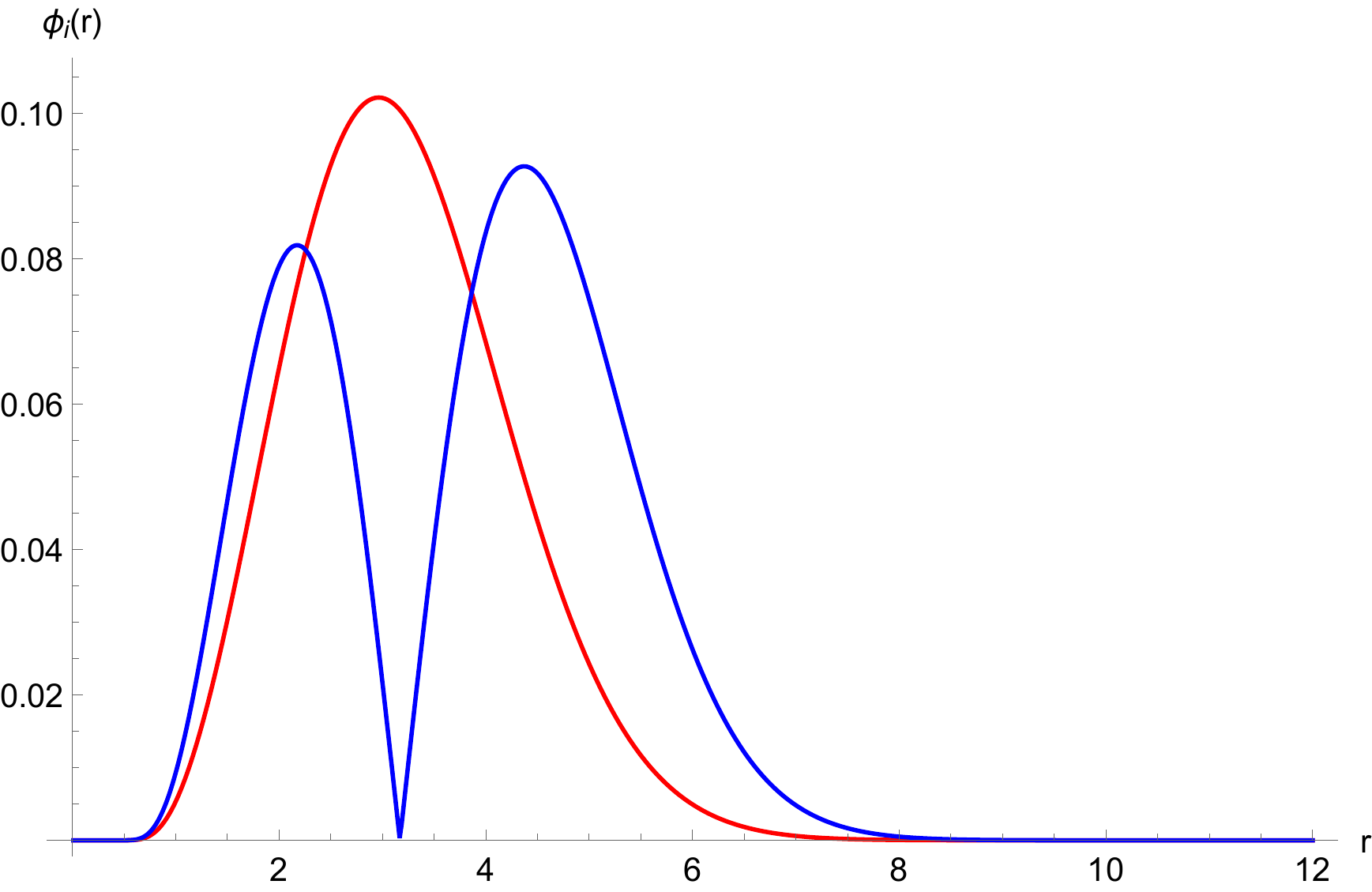}
\caption{The $l=0$ radial wave functions for single heavy-baryon (top) and double-heavy-baryon or penta-quark state (bottom).  See text. }\label{fig:Wavel=01}
\end{figure}

\section{Spectra}~\label{SPECTRA}


Given the Hamiltonian and the explicit induced quantum potentials,
we can now obtain  the spectra of the holographic heavy-light hadrons.
Our strategy is the following: we  treat the warping contribution as a small perturbation, while solving
the radial part numerically. For the warping part, using the average

\begin{align}
\label{SP34}
    \left<\frac{Z^2}{\rho^2}\right>_{n_z,n_\rho}=\frac{n_z+\frac{1}{2}}{\tilde l+1} \ ,
\end{align}
we obtain

\begin{align}
\delta M_{\rm wrap}=\frac{37+\frac{6}{\tilde l+1}}{192m_H} (N_Q+N_{\bar Q})  \ ,
\end{align}
in unit of $M_{KK}$.

To obtain the radial part, we need to solve the Schroedinger equation

\begin{align}
\label{SHRO1}
-\frac{1}{2m_y \rho^3(1+\frac{1}{2m_Hm_y\rho^2})^2}\frac{\partial}{\partial \rho}\left(\rho^3\bigg(1+\frac{1}{2m_Hm_y\rho^2}\bigg)\frac{\partial}{\partial \rho}\right)\Psi_{n,l}(\rho)+V_l(\rho)\Psi_{n,l} (\rho)=E_{n,l}\Psi_{n,l}(\rho) \ ,
\end{align}
with the warped  normalization condition

\begin{align}
2\pi^2\int_{0}^{\infty} \rho^3\bigg(1+\frac{1}{2m_y\rho^2}\bigg)^2 |\Psi_{n,l}(\rho)|^2d\rho=1 \ .
\end{align}
For this purpose we perform the transformation $\Psi \rightarrow u$ and use $\tilde \rho^2=m_y\rho^2$
\begin{align}
\Psi=\frac{u}{\sqrt{\rho^3(1+\frac{1}{2m_Hm_y\rho^2})}} \ ,
\end{align}
to simplify (\ref{SHRO1})
\begin{align}
-\frac{1}{2(1+\frac{1}{2m_H\tilde \rho^2})}u_{n,l}''(\tilde \rho)+\bigg[\frac{3\big(1+\frac{1}{m_H\rho^2}-\frac{1}{12m_H^2\tilde \rho^4}\big)}{8\tilde \rho^2(1+\frac{1}{2m_H\tilde \rho^2})}\bigg] u_{n,l}(\tilde \rho) +V_l(\tilde \rho)u_{n,l}(\tilde \rho)=E_{n,l} u_{n,l}(\tilde \rho) \ ,
\end{align}
with the normalization condition
\begin{align}
2\pi^2\int_{0}^{\infty} d\tilde \rho \bigg(1+\frac{1}{2m_H\tilde \rho^2}\bigg)|u_{n,l}(\tilde \rho)|^2=1 \ .
\end{align}
Notice that the normalization condition actually requires the $u_{n,l}$ to vanish near $\tilde \rho$.
In this case one can show that although the additional term ( large bracket in (\ref{SHRO1}) )  becomes negative at small $\tilde \rho$, the spectrum $E_{n,l}$ is still bounded from below. The above equation for $u_{n,l}$ can be diagonalized numerically and below we present the results for different states.

\begin{widetext}
\begin{table}[h]
\caption{Charm baryons  and Pentaquarks}
\begin{center}
\begin{tabular}{ccccccccc}
\hline
\hline
$B$ & $IJ^P$  &  $l$  & $n_\rho$ & $n_z$  & $N_Q$ &  $N_{\bar Q}$ &  Mass-MeV & Exp-MeV \\
\hline
\hline
$\Lambda_c$     &$0{\frac 12}^+$ & 0  & 0&  0& 1& 0&  2286 & 2286 \\
$\Sigma_c$        &$1{\frac 12}^+$ & 2  & 0&  0& 1& 0& 2557& 2453 \\
                           &$1{\frac 32}^+$ & 2  & 0&  0& 1& 0& 2596 & 2520 \\
$\Lambda^*_c$  &$0{\frac 12}^-$ & 0  & 0&  1& 1& 0& 2683 & 2595 \\
                           &$0{\frac 12}^+$ & 0  & 1&  0& 1& 0& 2726 & 2765\\
$\Sigma^*_c$     &$1{\frac 12}^-\,,\,1{\frac 32}^-$ & 2  & 0&  1& 1& 0& [2947/2986] & --\\
                           &$1{\frac 12}^+\,,\,1{\frac 32}^+$& 2 & 1  & 0&  1& 0&  [2948/2995] & --\\
$P_c$                 &$\frac 12{\frac 12}^-\,,\,\frac 12{\frac 32}^-$ & 1  & 0&  0& 1& 1&  [4340/4360/4374]& [4312/4440/4457] \\
$P^*_c$              &$1{\frac 12}^-\,,\,1{\frac 32}^-$ & 1  & 0&  1& 1& 1& [4732/4752/4767] & --\\
                           &$1{\frac 12}^+\,,\,1{\frac 32}^+$& 1 & 1  & 0&  1& 1& [4725/4746/4763] & --\\
\hline
\hline
\end{tabular}
\end{center}
\label{tab_bindtet}
\end{table}%
\end{widetext}

\begin{widetext}
\begin{table}[h]
\caption{Bottom baryons  and Pentaquarks}
\begin{center}
\begin{tabular}{ccccccccc}
\hline
\hline
$B$ & $IJ^P$  &  $l$  & $n_\rho$ & $n_z$  & $N_Q$ &  $N_{\bar Q}$ & Mass-MeV & Exp-MeV \\
\hline
\hline
$\Lambda_b$     &$0{\frac 12}^+$ & 0  & 0&  0& 1& 0&5608 & 5620 \\
$\Sigma_b$        &$1{\frac 12}^+$ & 2  & 0&  0& 1& 0& 5962 & 5810 \\
                           &$1{\frac 32}^+$ & 2  & 0&  0& 1& 0& 5978 & 5830\\
$\Lambda^*_b$  &$0{\frac 12}^-$ & 0  & 0&  1& 1& 0& 5998& 5912 \\
                           &$0{\frac 12}^+$ & 0  & 1&  0& 1& 0& 6029& (6072)\\
$\Sigma^*_b$     &$1{\frac 12}^-\,,\,1{\frac 32}^-$ & 2  & 0&  1& 1& 0& [6351/6367]& --\\
                           &$1{\frac 12}^+\,,\,1{\frac 32}^+$& 2 & 1  & 0&  1& 0&  [6344/6367] & --\\
$P_b$                 &$\frac 12{\frac 12}^-\,,\,\frac 12{\frac 32}^-$ & 1  & 0&  0& 1& 1& [11155/11163/11167]& --\\
$P^*_b$              &$1{\frac 12}^-\,,\,1{\frac 32}^-$ & 1  & 0&  1& 1& 1& [11544/11553/11556] & --\\
                           &$1{\frac 12}^+\,,\,1{\frac 32}^+$& 1 & 1  & 0&  1& 1& [11532 /11543/11579]& --\\
\hline
\hline
\end{tabular}
\end{center}
\label{tab_bindtetb}
\end{table}%
\end{widetext}


To fix the parameters for the charmed heavy baryons,  we choose $M_D=1.87$ GeV for the D-meson mass in (\ref{MDMH}) and fix $M_{KK}=0.475$ GeV
to reproduce the $M_{\Lambda_c}=2.286$ GeV. This low value of $M_{KK}$ is consistent with the value used to reproduce the nucleon spectra~\cite{Hata:2007mb},
but about half  the value of $M_{KK}\sim1$ GeV used originally in~\cite{Sakai:2004cn} and adopted in~\cite{Liu:2016iqo,Liu:2017xzo,Liu:2017frj}.
In this case we have  $m_H=(1.87-0.168)$ GeV= $3.66 M_{KK}$. In Fig.~\ref{fig:Wavel=01} we show the radial wavefunctions for the first and second excited states
following from (\ref{SHRO1}) for a single heavy-baryon (top) and doubly heavy-baryon or pentaquark state (bottom). Note the rapid decay of the wavefunctions near
the instanton core as $\rho\rightarrow 0$.

The corresponding  charm and bottom states for single- and double-heavy hadrons are listed in Table~\ref{tab_bindtet}
and Table~\ref{tab_bindtetb} respectively. Note that while $m_{\Lambda_c}=2.286$ GeV  is fitted to fix the Kaluza-Klein scale $M_{KK}=0.475$ GeV,
$m_{\Lambda_b}=5.608$ GeV is a holographic prediction which is remarkably close to the experimental value of 5.620 GeV.
The details of the mass budgets for each of the states in terms of the three  holographic parameters, are given in Appendix~\ref{DETAILS}.
The results for the single-heavy baryon spectrum are remarkable given the small number of parameters used in this holographic approach.
The spin contributions improve considerably the predictions for the masses and their hierarchy. In particular, the empirical mass ordering
$\Sigma_c-\Lambda_c<\Lambda_c^*-\Lambda_c$ is obtained contrary to the claim in~\cite{Fujii:2020jre}. The mass splitting between
$\Sigma_c$ and $\Sigma_b$ is higher than observed due to the sizable  repulsion from the $l=2$ intrinsic angular momentum assignment.

The holographic
construction with spin corrections, allows for only three pentaquark states which are  close to the observed charmed pentaquark states reported by LHCb,
although with slightly smaller masses (the 80 MeV difference can be easily narrowed by adjusting the Kaluza-Klein scale $M_{KK}=0.475$  GeV at the expense
of $\Lambda_c$).
The spin-orbit  effects split away the $[\frac 12\frac 12^-]_{S=1}$ and $[\frac 12\frac 32^-]_{S=1}$ states, lifting the degeneracy  reported
originally in~\cite{Liu:2017xzo}. The present holographic construction rules out a pentaquark with $[\frac 12\frac 52^\pm]$ assignment
since the instanton core carries equal spin-isospin~\cite{Liu:2017xzo}.   The splitting between the different pentaquark states are somehow
smaller than expected,  due to the strength of the spin-orbit coupling to order ${1}/{m_H^2}$. Additional contributions are expected to order
${1}/{m_H^3}$. This construction supports  additional  Roper-like and odd-parity-like pentaquark states
which we have denoted by $P^*_{c,b}$, although heavier and mor susceptible to decay.

\section{Conclusions}~\label{CONCLUSION}

In the holographic construction presented in~\cite{Liu:2016iqo,Liu:2017xzo,Liu:2017frj}, heavy hadrons are described in bulk
using a set of degenerate $N_f$ light D8-D$\bar 8$ branes  plus one heavy probe brane in the
cigar-shaped geometry that spontaneously breaks chiral symmetry. This construction enforces both chiral and
heavy-quark symmetry and describes well  the low-lying
heavy-light mesons and baryons. Heavy baryons are composed of heavy-light mesons bound to a core instanton in bulk.
Remarkably, the bound heavy-light  mesons with spin-1 transmute to heavy quarks with spin-$\frac 12$, an amazing spin-statistics transmutation
by geometry.

In~\cite{Liu:2016iqo,Liu:2017xzo,Liu:2017frj} the analysis of the bound states and spectra was carried to order $m_H^0$
where the spin effects are absent.
In this work and for $N_f=2$, we have now carried the analysis at next to leading order in $1/m_H$ where the spin-orbit and spin corrections are manifest.
By refining the Kaluza-Klein scale $M_{KK}$ from 1 GeV used in~\cite{Liu:2017xzo,Liu:2017frj}  to 0.475 GeV used here,
a rich spectrum with single- and double-heavy baryons emerges with fair  agreement with the empirically observed
states, including the newly reported charm pentaquark states by LHCb.

This is remarkable, given  that only three parameters were used in the holographic construction:
 $M_0, M_{KK}, m_H$.  For charm, they are fixed by
$M_0\rightarrow m_N$ (nucleon mass), $ M_{KK}\rightarrow M_{\Lambda_{c}}$ (Lambda-mass) and $m_H\rightarrow M_{D}$ (D-meson mass).
The only parameter adjustment for the bottom spectrum is $m_H\rightarrow M_B$ (B-meson mass).
Needless to say that the light-light,   heavy-light and heavy-heavy  mesons and baryons are described
simultaneously, without changing the number of parameters.

The holographic construction predicts a triplet of nearly degenerate charm pentaquark states  with the isospin-spin-parity assignments
$$\bigg(\bigg[P_c(4340)\frac 12\frac 12^-\bigg]_{S=1}, \bigg[P_c(4360)\frac 12\frac 12^-\bigg]_{S=0}, \bigg[P_c(4374)\frac 12\frac 32^-\bigg]_{S=1}\bigg)$$
which are to be compared to $P_c[4312|4440|4457]$ recently reported by LHCb. The small mass discrepancy can be readily eliminated by adjusting
the Kaluza-Klein scale at the expense of $\Lambda_c$.
The spin-orbit effects split away the states with intrinsice spin $S=1$.
The analysis rules out the assignment $[\frac 12\frac 52^\pm]$ for these states, and predicts a triplet of bottomed pentaquark states
$$\bigg(\bigg[P_b(11532)\frac 12\frac 12^-\bigg]_{S=1}, \bigg[P_b(11543)\frac 12\frac 12^-\bigg]_{S=0}, \bigg[P_b(11579)\frac 12\frac 32^-\bigg]_{S=1}\bigg)$$
 not yet observed. New Roper-like and odd-parity pentaquark states are also suggested, although much heaviers and more susceptible to fall apart.

Finally, the present holographic description can be regarded as  the holographic dual of the chiral soliton construction of heavy-light baryons~\cite{Rho:1992yy,Rho:MULTI}
(and references therein).
However, in the latter the uncertainties in  combining chiral and heavy quark symmetry strongly limit their predictive range, especially when
addressing the spin corrections. This is not the case for the holographic description as we have shown, as both symmetries are geometrically
embedded in the bulk brane construction with just three parameters. The dual approach is vastly superior.

\vskip 1cm
{\bf Acknowledgements}

This work is supported by the Office of Science, U.S. Department of Energy under Contract No. DE-FG-88ER40388 and  by the Polish National Science Centre (NCN) Grant UMO-2017/27/B/ST2/01139.

\appendix

\section{Details of the heavy mass expansion}~\label{HLACTION}

Following the rescaling in (\ref{S3}) the effective action for the heavy-light  fields split into the following contributions

\bea
\label{ACTIONALL}
{\cal L}=aN_c\lambda {\cal L}_0+aN_c({\cal L}_1+\tilde {\cal L}_1)+{\cal L}_{\rm CS}
\eea
with each contribution given by

\bea
{\cal L}_0=&&-(D_M\Phi_N^{\dagger}-D_N\Phi_M^{\dagger})(D_M\Phi_N-D_N\Phi_M)+2\Phi_M^{\dagger}F_{MN}\Phi_N  \ ,\nonumber \\
{\cal L}_1=&&+2(D_0\Phi_M^{\dagger}-D_M\Phi_0^{\dagger})(D_0\Phi_M-D_M\Phi_0)-2\Phi_0^{\dagger}F^{0M}\Phi_M\nonumber\\
&&-2\Phi^{\dagger}_MF^{M0}\Phi_0 -2m_H^2\Phi^{\dagger}_M\Phi_M \ ,
\nonumber\\
\tilde {\cal L}_1=&&+\frac{z^2}{3}(D_i\Phi_j-D_j\Phi_i)^{\dagger}(D_i\Phi_j-D_j\Phi_i) \nonumber \\
&&-2z^2(D_i\Phi_z-D_z\Phi_i)^{\dagger}(D_i\Phi_z-D_z\Phi_i)
-\frac{2}{3}z^2\Phi_i^{\dagger}F_{ij}\Phi_j+2z^2(\Phi^{\dagger}_zF_{zi}\Phi_i+c.c) \ \nonumber\\
{\cal L}_{CS}=&&-\frac{iN_c}{16\pi^2}\Phi^{\dagger}(dA+A^2)D\Phi-\frac{iN_c}{16\pi^2}(D\Phi)^{\dagger}(dA+A^2) \Phi +{\cal O}(\Phi^3)\ .
\eea
We now use the expansion  (\ref{SP1}) to explicitly derive  the various contributions in (\ref{ACTIONALL}) in leading order in $1/m_H$. The net result
has manifest heavy quark symmetry to order $m_H^0$, with the spin-orbit and spin-spin contributions  breaking this symmetry to order $1/m_H$.

\subsection{Kinetic contribution: ${\cal L}_{\rm kin}$}

 The explicit form of the kinetic contribution is

\begin{align}
\label{AP1}
{\cal L}_{\rm kin}=\frac{1}{8m_H}\chi^{\dagger}\sigma_M f(-\partial_t-\hat A_0-\Phi)(\partial_t+\hat A_0+\Phi)\bar \sigma_M f \chi \ ,
\end{align}
which contains derivative of $\chi$.  With the help of the identity for Weyl matrices  $\sigma_M \tau^a \bar \sigma_M=0$,  (\ref{AP1})  reads

\begin{align}
\label{AP2}
\frac{1}{2m_H}f^2\dot \chi^{\dagger}\dot \chi+ \frac{1}{2m_H} \dot X_N \dot X_M \partial_N f\partial_M f \chi^{\dagger}\chi
+\frac{f^2}{2m_H}\hat A_0 (\dot \chi^{\dagger}\chi-\chi^{\dagger}\dot \chi)
-\frac{f^2}{2m_H} \hat A_0^2 \chi^{\dagger}\chi-\frac{f^2}{4m_H} {\rm tr}(\Phi)^2 \chi^{\dagger}\chi \ .
\end{align}
which can be further simplified by using the explicit relations

\begin{align}
-{\rm tr} \Phi^2=\frac{X^4}{4(X^2+\rho^2)^2}2\chi^{a\dagger}\chi^a-\dot X_N \dot X_M {\rm tr}A_{N}A_{M}=\frac{2X^4}{(X^2+\rho^2)^2}\dot a_I^2+\frac{3X^2}{2(X^2+\rho^2)^2}\dot X^2
\end{align}
and

\begin{align}
    \hat A_0=-\frac{i}{8\pi^2 a x^2}\left(1-\frac{\rho^4}{(x^2+\rho^2)^2}\right) \ .
\end{align}
to have

\begin{align}\label{eq:Lkin}
    &{\cal L}_{\rm kin}\nonumber \\ &=\frac{\dot \chi^{\dagger}\dot \chi}{2m_H} + \left(\frac{1}{4m_H }\frac{\dot\rho^2}{\rho^2}+\frac{ \dot a_I^2}{4m_H}+\frac{\dot X^2}{4m_H \rho^2}\right)\chi^{\dagger}\chi+\frac{1}{16m_H\pi^2 a\rho^2}i\chi^{\dagger}\partial_t \chi+\frac{13}{3840m_H\pi^4a^2\rho^4}\chi^{\dagger}\chi
\end{align}
after integration over space.

\subsection{Chern-Simons contribution: ${\cal L}_{\rm CS}$}

The Chern-Simons term is

\begin{align}
{\cal L}_{\rm CS}&=-\frac{iN_c}{16\pi^2}\Phi^{\dagger}(dA+A^2)D\Phi-\frac{iN_c}{16\pi^2}(D\Phi)^{\dagger}(dA+A^2) \Phi \nonumber \\
&=-\frac{iN_c}{8\pi^2}\Phi^{\dagger}(dA+A^2)D\Phi
\end{align}
where in the second line we have performed a partial integration with the help of the Bianchi identity $DF=0$.  More explicitly, we have

\begin{align}\label{eq:CS}
&\frac{i}{128m_H\pi^2a}f^2\chi^{\dagger}\sigma_MF_{MN}\bar\sigma_N \dot \chi+\frac{i\epsilon_{MNPQ}}{128m_H\pi^2a}f\chi^{\dagger}\sigma_M F_{0N}(\partial_P+A_P)f\bar \sigma_{Q}\chi \nonumber \\
&+\hat A_0\left(\frac{iN_c}{128m_H\pi^2a}\chi^{\dagger} \sigma_M F_{MN}\bar \sigma_N \chi f^2\right) \ ,
\end{align}
which is seen to contain $\chi^{\dagger}\chi$ as well as linear terms in derivatives.  Recall that the electric field $F_{0M}$ after solving Gauss constraint reads

\begin{align}
F_{0M}=\dot X_N F_{MN}+\dot \rho \frac{\partial A_M}{\partial \rho}-\chi^a D_M\Phi^a \ ,
\end{align}
 The linear terms in $\dot \rho$, $\dot X_N$ vanish due to parity and translational invariance,  but there are terms of the form
\begin{align}
\chi^a \chi^{\dagger}\frac{\tau^a}{2}\chi=i\chi^{\dagger}{\bf a}^{-1}\dot{\bf a}\chi
\end{align}
which couple to isospin.  Again, using the identity for Weyl matrices

\begin{align}
\epsilon_{MNPQ}\bar \sigma_{QR}=-\delta_{MR}\bar \sigma_{MP}+\delta_{NR}\bar \sigma_{MP}-\delta_{PQ}\bar \sigma_{MN} \ ,
\end{align}
all terms that require anti-symmetrization vanish

\begin{align}
-\epsilon_{MNPQ}\sigma_M F_{0N}\bar \sigma_P \partial_Q f=0 \ ,
\end{align}
but the more involved one

\bea
-\epsilon_{MNPQ}\sigma_M F_{0N}A_P\bar \sigma_Q =&&\frac{\sigma_M F_{0N}\bar \sigma_{NQ}\bar \sigma_Q X_M}{X^2+\rho^2}-\frac{\sigma_M F_{0N}\bar \sigma_{MQ}\bar \sigma_QX_N}{X^2+\rho^2}\nonumber \\ &&+\frac{\sigma_M F_{0N}\bar \sigma_{MN}\bar \sigma_Q X_Q}{X^2+\rho^2} \ .
\eea
does not.
Using the identity $\sigma_M \tau^i \bar \sigma_M=0$, the second term vanishes, while the first and the third term read

\bea
\frac{\sigma_M F_{0N}\bar \sigma_{NQ}\bar \sigma_Q X_M}{X^2+\rho^2}=&&\frac{3\sigma \cdot X F_{0N}\bar \sigma_N}{X^2+\rho^2}=- \frac{9\rho^2f}{(X^2+\rho^2)^2} \tau^a \chi^a \ , \nonumber\\
\frac{\sigma_M F_{0N}\bar \sigma_{MN}\bar \sigma_Q X_Q}{X^2+\rho^2}=&&-\frac{\sigma_M F_{0N}\bar \sigma_N \sigma_M\bar \sigma \cdot X}{2(X^2+\rho^2)}= \frac{3\rho^2f}{(X^2+\rho^2)^2}\tau^a \chi^a \ ,
\eea
Since
\begin{align}
-\epsilon_{MNPQ}\sigma_M F_{0N}A_P\bar \sigma_Q= -\frac{6 \rho^2f}{(X^2+\rho^2)^2} \tau^a \chi^a  \ .
\end{align}
we finally have

\begin{align}
    -\epsilon_{MNPQ}\Phi^{\dagger}_M F_{0N}D_P\Phi_Q=-\frac{6 \rho^2f^2}{16m_HN_ca(X^2+\rho^2)^2}\chi^{\dagger}\tau^a\chi \chi^a \ .
\end{align}
One should also consider the contribution from $\hat F_{0N}=-\frac{1}{r}\frac{\partial {\hat A}_0}{\partial r} x_N$,

\begin{align}
    -\epsilon_{MNPQ}\sigma_M \hat F_{0N}A_P\bar \sigma_Q=\frac{6X^2}{X^2+\rho^2}\frac{1}{r}\frac{\partial \hat A_0}{\partial r} \,
\end{align}
The final Chern-Simons contribution to order $1/m_H$  after rescaling is

\begin{align}
   &\frac{3}{16m_H\pi^2a}\frac{f^2\rho^2}{(X^2+\rho^2)^2}i\chi^{\dagger}(\partial_t+\hat A_0)\chi+ \frac{3}{64m_H\pi^2a}\frac{\rho^2f^2}{(X^2+\rho^2)^2}\chi^{\dagger}\tau^a\chi i\chi^a \nonumber \\ &-\frac{3i}{64m_H \pi^2 a}\frac{X^2 f^2}{X^2+\rho^2}\frac{1}{r}\frac{\partial \hat A_0}{\partial r}\chi^{\dagger}\chi \ .
\end{align}
In fact, the first term can be obtained from the leading order result by noticing that $\partial_t\rightarrow -im_H+\partial_t$ and requiring gauge in variance. Using

\begin{align}
    \hat A_0=-\frac{i}{8\pi^2 a x^2}\left(1-\frac{\rho^4}{(x^2+\rho^2)^2}\right) \ ,
\end{align}
and  performing the spatial  integration we finally have

\begin{align}\label{eq:LCS}
    {\cal L}_{\rm CS}=\frac{1}{32m_H \pi^2a\rho^2}i\chi^{\dagger}\partial_t\chi+\frac{7}{1280m_H\pi^4a^2\rho^4}\chi^{\dagger}\chi+\frac{1}{128m_H\pi^2a \rho^2}\chi^{\dagger}\tau^a\chi i\chi^a
\end{align}

\subsection{The contribution:  $ \Phi_0$}

This is the most difficult term to unravel to order $1/m_H$. The equation of motion for $\Phi_0$ reads

\begin{align}
\label{AP10}
(-D_M^2+m_H^2) \Phi_0+2F_{M0}\Phi_M-\frac{i}{16\pi^2 a}F_{PQ}(\partial_P+A_P) \Phi_Q=0 \ .
\end{align}
after using the self-dual condition for $F$. Using  the standard relations for $\bar \sigma_{MN}$, we  have for the last two contributions
in (\ref{AP10})

\begin{align}
\label{AP11}
F_{PQ}\partial_P \Phi_Q=\frac{6\rho^2}{(X^2+\rho^2)^2} \frac{1}{r}\frac{ df}{dr}\bar\sigma \cdot X \chi \ ,\\
F_{PQ}A_P\Phi_Q=-\frac{6 \rho^2}{(X^2+\rho^2)^3}f\bar\sigma \cdot X\chi  \ .
\end{align}
For the first contribution in (\ref{AP10}) we have

\begin{align}
\label{AP12}
F_{M0}\Phi_M=\frac{6f}{(X^2+\rho^2)^2}\left(\rho^2 \bar \sigma \cdot \dot X+\bar \sigma \cdot X \rho\dot\rho\right)\chi+\chi^aD_M\Phi^a \bar \sigma_M \chi f \ .
\end{align}
with

\begin{align}
\Phi^a=\frac{1}{2(X^2+\rho^2)}\bar \sigma \cdot X \tau^a \sigma \cdot X \ ,
\end{align}
or more explicitly

\begin{align}
\label{AP11}
\chi^aD_M\Phi^a \bar \sigma_M \chi f=  \frac{3\rho^2 f}{(X^2+\rho^2)^2}\bar \sigma \cdot X \tau^a \chi \chi^a \ .
\end{align}
Inserting (\ref{AP11}-\ref{AP12}) into (\ref{AP10})  we  have

\begin{align}
\label{AP12}
(-D_M^2 +m_H^2)\Phi_0+J_0=0 \ ,
\end{align}
with

\begin{align}
J_0=&\frac{12f}{(X^2+\rho^2)^2}\left(\rho^2 \bar \sigma \cdot \dot X+\bar \sigma \cdot X \rho\dot\rho\right)\chi+ \frac{6f \rho^2}{(X^2+\rho^2)^2}\bar \sigma \cdot X \tau^a \chi \chi^a \nonumber \\
&+\frac{3i}{2\pi^2 a}\frac{\rho^2 f}{(X^2+\rho^2)^3} \bar \sigma \cdot X \chi + \frac{2f}{r}\frac{\partial \hat A_0}{\partial r} \bar \sigma \cdot X \chi \ ,
\end{align}
 the source for $\Phi_0$. In this equation the Abelian part of $F_{N0}$ has been included.  Since

\begin{align}
    \frac{1}{r}\frac{\partial \hat A_0}{\partial r}=\frac{i}{4\pi^2 a}\frac{1}{(X^2+\rho^2)^2}\left(1+\frac{2\rho^2}{X^2+\rho^2}\right)
\end{align}
one finally has

\begin{align}
   J_0=&\frac{12f}{(X^2+\rho^2)^2}\left(\rho^2 \bar \sigma \cdot \dot X+\bar \sigma \cdot X \rho\dot\rho\right)\chi+ \frac{6\rho^2f}{(X^2+\rho^2)^2}\bar \sigma \cdot X \tau^a \chi \chi^a \nonumber \\
&+\frac{i}{2\pi^2 a}\frac{f}{(X^2+\rho^2)^2}\left(1+\frac{5\rho^2}{X^2+\rho^2}\right) \bar \sigma \cdot X \chi
\end{align}

In the large $m_H$ limit, the contribution to $\Phi_0$ is ${1}/{m_H^2}$ suppressed compared to $\Phi_M$
\begin{align}
\Phi_0=-\frac{1}{-D_M^2+m_H^2}J_0=-\frac{1}{m_H^2}J_0+{\cal O}\left(\frac{1}{m_H^4}\right) \ ,
\end{align}
and can be neglected from the Lagrangian.

\subsection{$\Phi_0$ at $m_H=0$}

In the  opposite limit of $m_H=0$, it is instructive to see how the field $\Phi_0$  solves the constraint equation.
To solve  (\ref{AP12}), we  define the Green function

\begin{align}\label{eq:GREEN}
G(X,Y)\equiv \frac{1}{-D_M^2}(X,Y)=\frac{\rho^2+\bar\sigma \cdot X \sigma \cdot Y}{4\pi^2(X^2+\rho^2)^{\frac{1}{2}}(X-Y)^2(Y^2+\rho^2)^{\frac{1}{2}}}\ ,
\end{align}
in terms of which the solution can be written as

\begin{align}
\Phi_0(X)=-\int d^4Y G(X,Y)J_0(Y) \ .
\end{align}
To perform the integral one needs the following elementary integrals

\begin{align}\label{eq:Inte1}
&\int d^4 Y \frac{\rho^2+\bar \sigma \cdot X \sigma \cdot Y}{4\pi^2 (X-Y)^2(Y^2+\rho^2)^{1/2}} \frac{\bar \sigma \cdot Y}{(Y^2+\rho^2)^{n+3/2}}\equiv f_n(X^2,\rho^2)\bar \sigma \cdot X
\end{align}
with

\begin{align}\label{eq:fn}
&f_n(X^2,\rho^2)=\frac{\Gamma(n)}{4\Gamma(n+2)}\frac{1 }{X^2}\bigg[\left(2-(n-2)\frac{\rho^2}{X^2}\right)\frac{\Delta(n-1,X^2)}{n-1}+\frac{1}{X^2}\Delta(n-2,X^2)\bigg] \ ,
\end{align}
and

\begin{align}\label{eq:Inte2}
&\int d^4 Y \frac{\rho^2+\bar \sigma \cdot X \sigma \cdot Y}{4\pi^2(X-Y)^2(Y^2+\rho^2)^{1/2}} \frac{1}{(Y^2+\rho^2)^{n+3/2}} \equiv g_n(X^2,\rho^2) \ ,
\end{align}
with

\begin{align}
  g_n(X^2,\rho^2)=\frac{1}{4(n+1)X^2}\frac{\Delta(n-1,X^2)}{n-1} \ .
\end{align}
Here  $\Delta(n,X^2)$ reads

\begin{align}\label{eq:gn}
    \Delta(n,X^2)=\frac{1}{\rho^{2n}}-\frac{1}{(X^2+\rho^2)^n} \ ,
\end{align}
with the  limit

\begin{align}
    \lim_{n\rightarrow 0}\frac{\Delta(n,X^2)}{n}=\ln (1+\frac{X^2}{\rho^2}) \ .
\end{align}
subsumed.
As $X\rightarrow 0$, $f_n$ and $g_n$ are all regular.
With the above in mind,  the explicit solution  for  $\Phi_0$  follows

\begin{align}\label{eq:phsolve}
    \Phi_0=-\frac{c}{(X^2+\rho^2)^{\frac{1}{2}}}\bigg[&12\rho^2 g_2(X^2,\rho^2) \bar \sigma \cdot \dot X+6\rho^2 f_2(X^2,\rho^2) \bar \sigma \cdot X \tau^a \chi^a\nonumber \\
    &+12f_2(X^2,\rho^2)\rho\dot \rho\bar \sigma \cdot X+\frac{i}{2\pi^2 a}\left(f_2(X^2,\rho^2)+5\rho^2 f_3(X^2,\rho^2)\right)\bar \sigma \cdot X\bigg]\chi \ .
\end{align}
where we have used the zero-mode profile

\begin{align}
    f=\frac{c}{(X^2+\rho^2)^{\frac{3}{2}}} \ .
\end{align}
with  $c=\sqrt{2}\rho/\pi$.

In terms of~(\ref{eq:phsolve}),  the $\Phi_0$
 contribution to the Lagrangian is

\begin{align}
 S= \frac{1}{8m_H}\int d^4X J_0^{\dagger}(X)\Phi_0(X) \ .
\end{align}
Using the fact that $\chi^a$ is anti-hermitian, all the mixing terms vanish, with the exception of

\begin{align}
   \frac{6i}{8\pi^2am_H}\int d^4 X \frac{c^2 \rho^2 X^2 }{(X^2+\rho^2)^{4}}\left(1+\frac{5\rho^2}{X^2+\rho^2}\right) f_2(X^2,\rho^2) \chi^{\dagger}\tau^a \chi \chi^a
\end{align}
which couples the spin of the nucleon core and the heavy-quarks.  After the spatial integration, it reads

\begin{align}
 \label{AP15}
    \frac{i}{32m_H\pi^2 a \rho^2}\chi^{\dagger}\tau^a \chi \chi^a  \ .
\end{align}
The diagonal terms  give

\begin{align}
    &-\frac{1}{8m_H}\bigg[\dot X^2 \chi^{\dagger}\chi \int d^4 X\frac{12^2\rho^4 c^2 g_2(X^2,\rho^2) }{(X^2+\rho^2)^{4}} +\rho^2\dot \rho^2 \chi^{\dagger}\chi \int d^4 X \frac{12^2c^2X^2f_2(X^2,\rho^2)}{(X^2+\rho^2)^{4}} \nonumber \\
    &+\chi^{a\dagger}\chi^a \chi^{\dagger}\chi \int d^4 X \frac{6^2c^2\rho^2X^2f_2(X^2,\rho^2) }{(X^2+\rho^2)^{4}}\nonumber \\ &+\frac{c^2\rho^4}{4\pi^4a^2}\chi^{\dagger}\chi\int d^4X \frac{X^2\left[f_2(X^2,\rho^2)+5\rho^2f_3(X^2,\rho^2)\right]}{(X^2+\rho^2)^4}\bigg(1+\frac{5\rho^2}{X^2+\rho^2}\bigg) \bigg]
\end{align}
and reduce to

\begin{align}
\label{AP16}
    -\chi^{\dagger}\chi\bigg(\frac{\dot X^2}{4m_H\rho^2}+\frac{1}{4m_H}\frac{\dot \rho^2 }{\rho^2}+\frac{1}{4 m_H}\dot a_I^2+\frac{25}{6144m_H\pi^4a^2\rho^4}\bigg)
\end{align}
after integration. (\ref{AP15}-\ref{AP16}) yield the final $\Phi_o$ contribution to the action

\begin{align}\label{eq:LPhi0}
    &{\cal L}_{\Phi_0}=\nonumber \\
    &-\chi^{\dagger}\chi\bigg(\frac{\dot X^2}{4m_H\rho^2}+\frac{1}{4m_H}\frac{\dot \rho^2 }{\rho^2}+\frac{1}{4 m_H}\dot a_I^2+\frac{25}{6144m_H\pi^4a^2\rho^4}\bigg)+\frac{i}{32m_H\pi^2 a \rho^2}\chi^{\dagger}\tau^a \chi \chi^a  \ .
\end{align}

\subsection{The warping contribution:  $\delta L_{\rm warp}$. }

The warping contribution stems from $\tilde{S}_1$ and does not have any derivative coupling. More specifically, we have

\bea
\label{AP17}
\delta \tilde S_1=&&\frac{(z+Z)^2}{24m_H}\bigg(\bigg(3-\frac{2z^2}{z^2+x^2}\bigg)f^{\prime 2}+\frac{6x^2+12z^2}{(x^2+z^2+\rho^2)^2}f^2\bigg)\chi^{\dagger}\chi\nonumber \\
&&-\frac{(z+Z)^2}{8m_H} \bigg((1+\frac{2z^2}{x^2+z^2})f^{\prime 2}+\frac{9x^2+3z^2}{(x^2+z^2+\rho^2)^2}f^2\bigg)\chi^{\dagger}\chi+\frac{\rho^2(z+Z)^2}{m_H(z^2+x^2+\rho^2)^2}f^2\chi^\dagger\chi \ .\nonumber\\
\eea
After  spatial integration, (\ref{AP17})  gives rise to  a $\frac{Z^2}{\rho^2} \chi^{\dagger} \chi$ term as well as a $\chi^{\dagger}\chi$ term,  namely

\begin{align}\label{eq:finalwrap}
    {\cal L}_{\rm warp}=-\frac{37+12\frac{Z^2}{\rho^2}}{192m_H}\chi^{\dagger}\chi \ .
\end{align}
 Notice that the  $Z^2$contribution  is negative, which is  consistent with an instability at large $Z$.

 \subsection{The contribution: ${\cal L}_0$}

 To leading order in $\lambda$, this contribution vanishes since $\Phi_M$ satisfies the equation of motion. However, there are contributions to $\hat A_M$ at order $1/\lambda$,

 \begin{align}
     {\cal L}_0=4aN_c\lambda \Phi^{\dagger}_M\Phi_N \hat F_{MN}=8aN_c\lambda \Phi^{\dagger}_M\Phi_N\partial_M\hat A_N \ .
 \end{align}
 To linear order in $\chi^a$, we need the explicit solution to $\hat A_M$

 \begin{align}
     \hat A_M=\frac{i}{16\pi^2a\lambda}\frac{\chi^a{\rm tr}\tau^a \sigma_{MN}X_N}{2(X^2+\rho^2)^2} \ .
 \end{align}
With this in mind and   using the  identities

 \begin{align}
     \sigma_{NM}=i\bar \eta^a_{NM}\tau^a \ , \\
     \bar \eta^a_{NM}\bar \eta^b_{NM}=4\delta^{ab} \ ,
 \end{align}
 we have

 \begin{align}
    8aN_c\lambda \Phi^{\dagger}_M\Phi_N\partial_M\hat A_N=\frac{1}{8m_H\pi^2a}\frac{\rho^4f^2}{(X^2+\rho^2)^3} i\chi^a\chi^{\dagger}\tau^a\chi \ ,
 \end{align}
 which  after spatial-integration reduces to
 \begin{align}
 \label{L0X}
     {\cal L}_0=\frac{i}{80m_H\pi^2a}\chi^a\chi^{\dagger}\tau^a\chi \ .
 \end{align}

\section{Coulomb-back reaction}
\label{BACK}

Here we provide a complete treatment of the Coulomb back interaction contribution. After re-scaling $A_0\rightarrow iA_0$, the Lagrangian for $A^0$ reads
\begin{align}
{\cal L}[A_0]=\frac{aN_c}{2}(\vec{\nabla} A_0)^2+\frac{f^2}{2m_H}\chi^{\dagger}\chi A_0^2+A_0(\rho^{cl}+\rho_0+\frac{1}{m_H}\rho_1)
\end{align}
where $\rho^{cl}$ is the source without the heavy-quark field
\begin{align}
\rho_c=aN_c\nabla^2 A^{cl}_0=-\frac{3N_c}{\pi^2}\frac{\rho^4}{(x^2+\rho^2)^4}
\end{align}
and we have
\begin{align}
\rho_0=&f^2\chi^{\dagger}\chi \ , \nonumber \\
\rho_1=&\frac{f^2}{2}i(\chi^{\dagger}\dot \chi-\dot\chi^{\dagger}\chi)+\frac{3}{16m_H\pi^2a}\frac{2\rho^2-X^2}{(X^2+\rho^2)^2} f^2 \chi^{\dagger}\chi \ .
\end{align}
Notice that
\begin{align}
\frac{3}{16m_H\pi^2a}\frac{2\rho^2-X^2}{(X^2+\rho^2)^2}
=\frac{3}{16m_H\pi^2a}\frac{f^2\rho^2}{(X^2+\rho^2)^2}\chi^{\dagger}\chi+\frac{3}{64m_H\pi^2a}\partial_N\left(\frac{x_Nf^2}{(x^2+\rho^2)}\right)\chi^{\dagger}\chi
\end{align}
originates purely from the Chern-Simions contribution.   Given the action for $A_0$, at  the minimum we have
\begin{align}
{\cal L}_{\rm coulumb}=-\left(\rho^{cl}+\rho_0+\frac{1}{m_H}\rho_1\right)\frac{1}{2\left(-aN_c \nabla^2+\frac{f^2}{m_H}\chi^{\dagger}\chi\right)}\left(\rho^{cl}+\rho_0+\frac{1}{m_H}\rho_1\right) \ ,
\end{align}
which is a complicated function in $\chi^{\dagger}\chi$ and always leads to positive energy.  In fact, the $\frac{f^2}{m_H}$ term in the denominator plays the role of a screening mass which can be seen after certain coordinate transformation.

\begin{figure}[h]
\includegraphics[width=0.7\columnwidth]{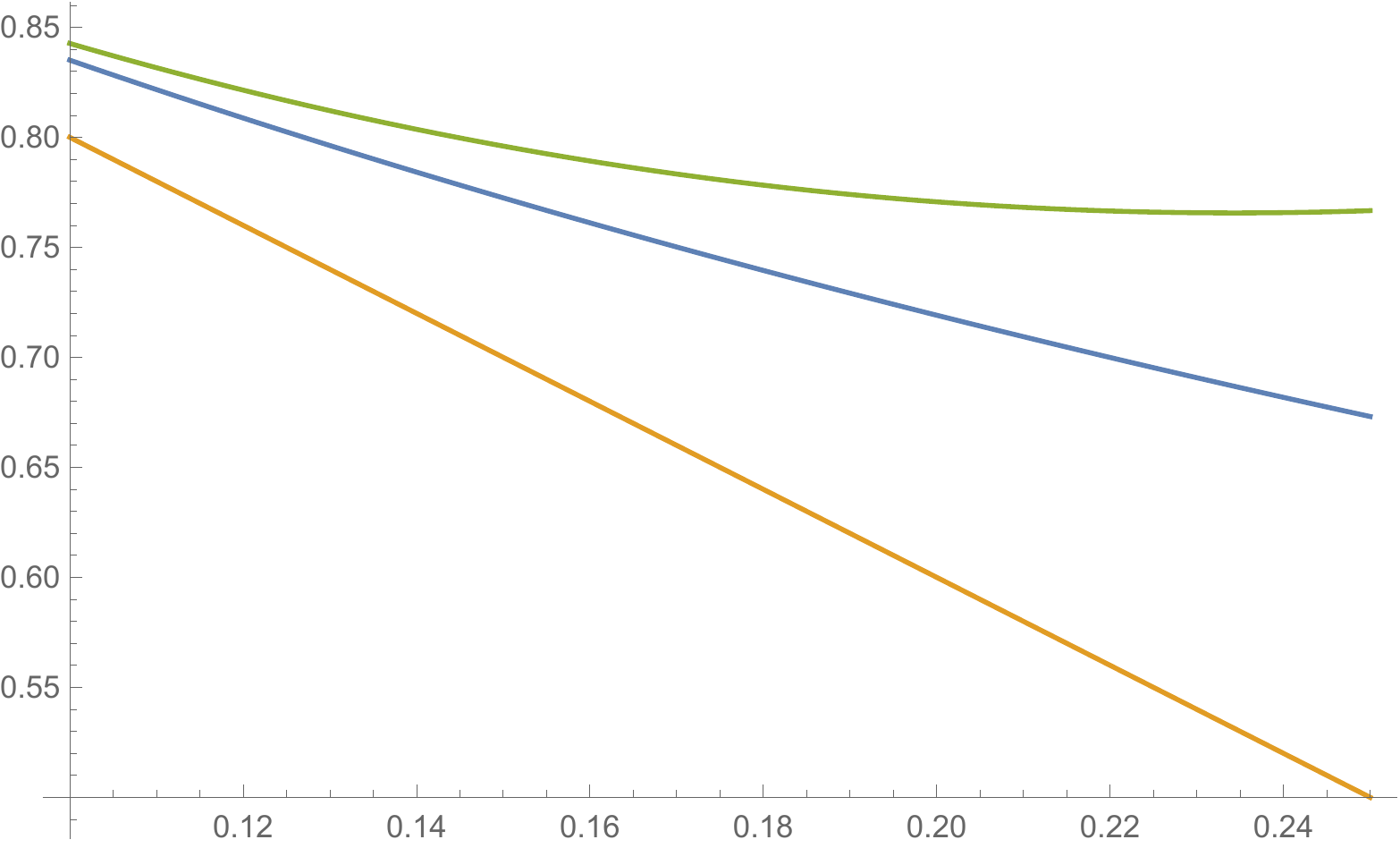}
\caption{The ratio $R(b\equiv\frac{32}{m_H})$ (blue) compared with its first order (yellow) and second order (green) Taylor expansion. At $\frac{1}{m_H}=\frac{1}{4}$ one has $R(b) \approx 0.67$, while at $\frac{1}{m_H}=\frac{1}{10}$ one has $R(b)\approx 0.84$. As $b\rightarrow \infty$, $R(b)\rightarrow 0$.  At $\frac{1}{m_H}=\frac{1}{4}$ the second order result is about $10\%$  larger than of the full result. }\label{fig:Rb}
\end{figure}

To estimate how good the first order expansion is, one can consider the simplest case where the inversion is acting only on the $\rho_0 \propto f^2$. To keep track of
the dependence on $\rho$ and $m_H$, it is useful  to perform the re-scaling
\begin{align}
X\rightarrow \frac{1}{\sqrt m_y} \tilde \rho \tilde X  \ , \qquad\qquad
\rho \rightarrow \frac{1}{\sqrt m_y} \tilde \rho \ ,
\end{align}
As a result we have
\begin{align}
\frac{1}{\left(-aN_c \nabla^2+\frac{f^2}{m_H}\chi^{\dagger}\chi\right)} f^2=\frac{32}{\tilde \rho^2}\frac{1}{-\tilde \nabla^2+ \frac{32\chi^{\dagger}\chi}{m_H \tilde \rho^2}\frac{1}{(\tilde X^2+1)^3}}\frac{1}{(\tilde X^2+1)^3} \ .
\end{align}
which can be exactly solved as
\begin{align}
\frac{1}{-\tilde \nabla^2+ \frac{32\chi^{\dagger}\chi}{m_H \tilde \rho^2}\frac{1}{(\tilde X^2+1)^3}}\frac{1}{(\tilde X^2+1)^3}=\frac{1}{b}-\frac{\sqrt{\tilde X^2(1+\tilde X^2)}I_1(\sqrt{\frac{b\tilde X^2}{1+\tilde X^2}})}{bI_1(\sqrt{b})\tilde X^2} \ ,
\end{align}
with $b=\frac{32\chi^{\dagger}\chi}{m_H \tilde \rho}$ .  Therefore, one has
\begin{align}
&f^2 \frac{1}{\left(-aN_c \nabla^2+\frac{f^2}{m_H}\chi^{\dagger}\chi\right)} f^2 \nonumber \\
&=\frac{64}{\pi^2\tilde \rho^2}\int d^4 \tilde X \frac{1}{(\tilde X^2+1)^3}\bigg(\frac{1}{b}-\frac{\sqrt{\tilde X^2(1+\tilde X^2)}I_1(\sqrt{\frac{b\tilde X^2}{1+\tilde X^2}})}{bI_1(\sqrt{b})\tilde X^2} \bigg) \ .
\end{align}
Notice that although the $\frac{1}{b}$ appears to be at variance with power-counting, the  Taylor expansion
\begin{align}
&g(b,\tilde X)\equiv \frac{1}{b}-\frac{\sqrt{\tilde X^2(1+\tilde X^2)}I_1(\sqrt{\frac{b\tilde X^2}{1+\tilde X^2}})}{bI_1(\sqrt{b})\tilde X^2}\nonumber \\
&=\frac{1}{8(\tilde X^2+1)}+ \left(-\frac{\tilde X^4}{192 \left(\tilde X^2+1\right)^2}+\frac{\tilde X^2}{64 \left(\tilde X^2+1\right)}-\frac{1}{96}\right)b\nonumber \\
&+\left(-\frac{\tilde X^6}{9216 \left(\tilde X^2+1\right)^3}+\frac{\tilde X^4}{1536 \left(\tilde X^2+1\right)^2}-\frac{\tilde X^2}{768 \left(\tilde X^2+1\right)}+\frac{7}{9216}\right)b^2+{\cal O}(b^3) \ ,
\end{align}
formally converges for any $b$. However, for the case where $\tilde \rho=1$ and $\chi^{\dagger}\chi=1$, one has $b=\frac{32}{m_H} \approx 8$ for charm and $\approx 3.2$ for bottom, the convergence is poor  for the first few terms. To perform an estimate, one can consider the ratio
\begin{align}
R(b)=\frac{\int d^4 \tilde X \frac{g(b,\tilde X)}{(1+\tilde X)^3}}{\int d^4 \tilde X \frac{\lim_{b \rightarrow 0}g(b,\tilde X)}{(1+\tilde X^2)^3}} \ .
\end{align}
which is shown in  Fig.~\ref{fig:Rb}. One can actually show that $R(b)$ is always positive and goes to zero as $b\rightarrow \infty$ or $\tilde \rho \rightarrow 0$, which implies a weaker repulsion compared to the Leading order Coulomb one. However, expanding to leading order in $b$, the potential becomes unbounded from below at large $b$ or small $\rho$. Apparently, this instability is caused by the breakdown of the small $b$ expansion near the core. To fix the instability, we can include the second order term in the expansion. In fact, in Fig.~\ref{fig:Rb} we note that after including the second-order term, the difference between the full result is around $10\%$ for ${1}/{m_H} \approx 4$ at $\tilde \rho \approx 1$ for the charm quark. It is even better for the bottom quark.

Using the explicit form of the inversion
\begin{align}
&\frac{1}{-\tilde \nabla^2+\frac{b}{(\tilde X^2+1)^3}}\frac{1}{(\tilde X^2+1)^4}\nonumber \\
&=\frac{\tilde X^2+2}{24 \left(X^2+1\right)^2}-\frac{\left(3 \tilde X^4+9 \tilde X^2+7\right)}{1152 \left(\tilde X^2+1\right)^3}b+\frac{ \left(16 \tilde X^6+64 \tilde X^4+86 \tilde X^2+39\right)}{92160 \left(\tilde X^2+1\right)^4}b^2\nonumber \\
&-\frac{\left(130 \tilde X^8+650 \tilde X^6+1220 \tilde X^4+1020 \tilde X^2+321\right)}{11059200 \left(\tilde X^2+1\right)^5}b^3\nonumber \\
&+\frac{\left(1485 \tilde X^{10}+8910 \tilde X^8+21365 \tilde X^6+25605 \tilde X^4+15345 \tilde X^2+3681\right)}{1857945600 \left(\tilde X^2+1\right)^6}b^4 \ ,
\end{align}
and
\begin{align}
&\frac{1}{-\tilde \nabla^2+\frac{b}{(\tilde X^2+1)^3}}\frac{1}{(\tilde X^2+1)^5} \nonumber \\
&=\frac{\tilde X^4+3 \tilde X^2+3}{48 \left(\tilde X^2+1\right)^3}-\frac{\left(18 \tilde X^6+72 \tilde X^4+98 \tilde X^2+47\right)}{11520 \left(\tilde X^2+1\right)^4}b +\frac{\left(50 \tilde X^8+250 \tilde X^6+470 \tilde X^4+395 \tilde X^2+126\right)}{460800 \left(\tilde X^2+1\right)^5}b^2 \nonumber \\
&-\frac{\left(575 \tilde X^{10}+3450 \tilde X^8+8275 \tilde X^6+9925 \tilde X^4+5958 \tilde X^2+1434\right)}{77414400 \left(\tilde X^2+1\right)^6}b^3 \nonumber \\
&+\frac{\left(26355 \tilde X^{12}+184485 \tilde X^{10}+537355 \tilde X^8+833875 \tilde X^6+727335 \tilde X^4+338205 \tilde X^2+65523\right)}{52022476800 \left(\tilde X^2+1\right)^7}b^4 \ .
\end{align}
(\ref{SP15}) follows to order  ${\cal O}({1}/{m_H^2})$.

\section{Details of the heavy pentaquark masses }
\label{DETAILS}

Here we detail the various contributions to the mass spectra recorded in Table~\ref{tab_bindtet}  and  Table~\ref{tab_bindtetb}.
For completeness, we recall  that we fix $M_D=1.87$ GeV to reproduce the D-meson mass in (\ref{MDMH}) and fix $M_{KK}=0.475$ GeV
to reproduce the $M_{\Lambda_c}=2.286$ GeV. As result, we have for the charmed heavy-light hadrons recorded in Table~\ref{tab_bindtet}

\begin{align}
&    M_{\Lambda_c}=m_H+m_N-0.82M_{KK}+0.253\frac{M_{KK}^2}{m_H}=2.286 {\text GeV} \ ,\\
 &   M_{\Sigma_c}(\frac{1}{2})=m_H+m_N-0.234M_{KK}+0.203\frac{M_{KK}^2}{m_H}=2.557{\text GeV} \ , \\
&    M_{\Sigma_c}(\frac{3}{2})=m_H+m_N-0.154M_{KK}+0.203\frac{M_{KK}^2}{m_H}=2.596{\text GeV} \ , \\
   & M_{\Lambda_c^{\star}}(P=-1)=m_H+m_N-0.82M_{KK}+\frac{2}{\sqrt{6}}M_{KK}+0.321 \frac{M_{KK}^2}{m_H}= 2.683{\text GeV} \ , \\
   & M_{\Lambda_c^{\star}}(P=1)=m_H+m_N+0.107M_{KK}+0.253\frac{M_{KK}^2}{m_H}=2.726{\text GeV} \ , \\
    &P_c(J=\frac{1}{2},S=0)=2m_H+m_N-0.078M_{KK}+0.404\frac{M_{KK}^2}{m_H}=4.360{\text GeV} \ , \\
    &P_c(J=\frac{1}{2},S=1)=2m_H+m_N-0.119M_{KK}+0.404\frac{M_{KK}^2}{m_H}=4.341{\text GeV} \ , \\
    &P_c(J=\frac{3}{2},S=1)=2m_H+m_N-0.05M_{KK}+0.404\frac{M_{KK}^2}{m_H}=4.373{\text GeV} \ .
\end{align}
For the bottom heavy-light hadrons we fix the heavy-light meson mass
$m_H=(5.28-0.168)$ GeV=$10.76 M_{KK}$.  The bottom heavy-light mass spectra recorded in Table~\ref{tab_bindtetb} follow from

\begin{align}
   & M_{\Lambda_b}=m_H+m_N-0.958M_{KK}+0.253\frac{M_{KK}^2}{m_H}=5.608 {\text GeV} \ ,\\
    &M_{\Sigma_b}(\frac{1}{2})=m_H+m_N-0.207M_{KK}+0.203\frac{M_{KK}^2}{m_H}=5.962{\text GeV} \ , \\
    &M_{\Sigma_b}(\frac{3}{2})=m_H+m_N-0.174M_{KK}+0.203\frac{M_{KK}^2}{m_H}=5.978{\text GeV} \ , \\
    &M_{\Lambda_b^{\star}}(P=-1)=m_H+m_N-0.958M_{KK}+\frac{2}{\sqrt{6}}M_{KK}+0.321 \frac{M_{KK}^2}{m_H}=5.998{\text GeV} \ ,\\
    &M_{\Lambda_b}^{\star}(P=1)=m_H+m_N-0.072M_{KK}+0.253\frac{M_{KK}^2}{m_H}=6.029{\text GeV} \ , \\
    &P_b(J=\frac{1}{2},S=0)=2m_H+m_N-0.0393M_{KK}+0.404\frac{M_{KK}^2}{m_H}=11.163{\text GeV} \ , \\
    &P_b(J=\frac{1}{2},S=1)=2m_H+m_N-0.056M_{KK}+0.404\frac{M_{KK}^2}{m_H}=11.155{\text GeV} \ , \\
    &P_b(J=\frac{3}{2},S=1)=2m_H+m_N-0.030M_{KK}+0.404\frac{M_{KK}^2}{m_H}=11.196{\text GeV} \ .
\end{align}

\bibliography{HL}

\end{document}